\documentclass[twoside]{article}
\usepackage{palatino,epsfig,latexsym,apalike}
\usepackage[latin1]{inputenc}
\usepackage[figuresright]{rotfloat}

\begin{document}

\title{\bf The Complex Network of Evolutionary Computation Authors:
an Initial Study}

\author{{\bf Carlos Cotta}\\
	{\sf ccottap@lcc.uma.es}\\
        Dept. Lenguajes y Ciencias de la Computación,\\
	University of Málaga, Spain, \\ \\
        {\bf Juan-Julián Merelo} \\ 
	{\sf jj@merelo.net}\\
        Dept. Arquitectura y Tecnología de Computadores, \\
	University of Granada, Spain }

\maketitle

\begin{abstract}
EC paper authors form a complex network of co-authorship which is,
by itself, an example of an evolving system with its own rules,
concept of fitness, and patterns of attachment. In this paper we
explore the network of authors of evolutionary computation papers
found in a major bibliographic database. We examine its macroscopic
properties, and compare it with other co-authorship networks; the EC
co-authorship network yields results in the same ballpark as other
networks, but exhibits some distinctive patterns in terms of
internal cohesion. We also try to find some hints on what makes an
author a sociometric star. Finally, the role of proceeding
editorship as the origin of long-range links in the co-authorship
network is studied as well.
\end{abstract}

\section{Introduction}

The study of all kind of networks has undergone an accelerated
expansion in the last few years, after the introduction of models
for power-law \cite{emerg:barabasi1999} and scale-free networks
\cite{watt98}, which, in turn, has induced the study of many
different phenomena under this new light. One of them have been
co-authorship networks: nodes in these networks are paper authors,
joined by edges if they have written at least a paper in common.
Even as most papers are written by a few authors staying at the same
institution, science is a global business nowadays, and lots of
papers are co-authored by scientists continents apart from each
other. There are several interesting facts that can be computed on
these co-authorship networks: first, what kind of macroscopic values
they yield, and second, which are the most outstanding \emph{actors}
(authors) and edges (co-authorships) within this network. A better
understanding of the structure of the network and what makes some
nodes stand out goes beyond mere curiosity to give us some insight
on the deep workings of science, what makes an author popular, or
some co-authors preferred over others.

Co-authorship networks are studied within the field of sociometry,
and, in the case at hand, scientometry. First studies date back to
the second half of the nineties: Kretschmer \cite{kretschmer97}
studied the {\em invisible colleges} of physics, finding that their
behavior was not much different to other collaboration networks,
such as co-starring networks in movies. However, it was at the
beginning of this century when Newman \cite{newman01a,newman01b}
studied co-authorship networks as complex networks, giving the first
estimations of their overall shape and macroscopic properties. In
general, these kind of networks are both small worlds \cite{watt98},
that is, there is, on average, a short distance between any two
scientists taken at random, and scale free, which means they follow
a power law \cite{emerg:barabasi1999} in several node properties
(e.g., the in-degree, or number of nodes linking a particular one) .
Newman made measurements on networks from several disciplines:
physics, medicine and computer science, showing results for
clustering coefficients (related to transitivity in co-authorship
networks), and mean and maximum distances (which gives an idea of
the shape of the network). Barab\'asi and collaborators
\cite{barabasi02} later proved that the scale free structure of
these co-authorship networks can be attributed to preferential
attachment: authors that have been more time in business publish
more papers on average, and thus get more new links than new
authors. However, even as this model satisfactorily explains the
overall structure of the network, there must be much more in the
author positions in the network than just having been there for more
time. In addition to these general works, several studies have also
focused in particular scientific communities: computer support of
cooperative work \cite{horn2004}, psychology and philosophy
\cite{psy2003}, chemistry \cite{chem2004}, SIGMOD authors
\cite{SIGMOD2003} and sociology \cite{moody04}, to name a few.

In this work, we analyze the co-authorship network of evolutionary
computation researchers. Studying this network gives us a better
understanding of its cohesiveness as a discipline, and sheds some
light on the collaboration patterns of the community. It also
provides interesting hints about who are the central actors in the
network, and what determines their prominency in the area.

\section{Materials and Methods}
\label{sec:mm}

The bibliographical data used for the construction of the
scientific-collaboration network in EC has been gathered from the
DBLP\footnote{\texttt{http://www.informatik.uni-trier.de/$\sim$ley/db/}}
--\emph{Digital Bibliography \& Library Project}-- computer Science
bibliography server, maintained by Michael Ley at the University of
Trier. This database provides bibliographic information on major
computer science journals and proceedings, comprising more than
610,000 articles and several thousand computer scientists (as of
March 2005).

The database provides bibliographical data indexed by author and by
conference/journal. This turns out to be one of its advantages
since, for example, the URL of the page containing the information
for a certain author can be used as identifying key for that author.
To some extent this alleviates one of the problems typically found
in this kind of studies, namely the fact that a single author may
report his/her name differently on different papers (e.g., using the
first name or just initials, including a middle name or not,
etc.)\footnote{A second kind of problem is possible: having two
different authors with exactly the same name. We are not aware of
any glaring instance of this duplicity in the EC community.}. Of
course, this kind of situation is still possible in this database,
and indeed we have found some instances of it. However, it seems
that the maintainers of the database have put some care in avoiding
this issue.

Besides this indexing issue, the DBLP exhibits two additional
advantages. Firstly, it is a ``moderated'' database, meaning that it
is not updated via authors' submitting their references. On the
contrary, the maintainers add themselves new entries by inspecting
published volumes, or incorporate full Bib\TeX\ collections provided
by publishers or editors. This eliminates a potential source of bias
in the sample of publications, i.e., some authors being very active
in submitting their bibliographical entries while other being less
proactive in this sense. Finally, the second additional advantage is
the fact that DBLP pages are highly structured and regular. Hence
they are very amenable for automated parsing by a scraping program.
In particular, hyperlinks are provided for every co-author of a
paper, making  navigation through the database very easy.

The process to obtain the raw data is the following: our scraping
robot is firstly fed with a collection of DBLP author keys, stored
in a stack. Subsequently, while this stack is not empty, a key is
extracted from it, and the corresponding HTML page is downloaded.
Then, it is parsed to extract the textual name of the author, and
the papers he/she has authored. For each of these papers, the
hyperlinks of co-authors are identified, and added to the stack
(cycles are avoided by keeping track of processed authors using an
ordered binary tree). An important issue to be taken into account is
the fact that we are interested in obtaining a network for the EC
community. However, an EC author may also publish articles in other
fields; hence, we cannot blindly parse all entries in a certain page
since non-EC papers (and later on, non-EC authors) would be included
in the network. To avoid this, we have used a double check: firstly
we look for certain patterns in the publication reference. These
include the acronyms of EC-specific conferences --such as GECCO,
PPSN, EuroGP, etc.-- or keywords --such as ``Evolutionary
Computation'', ``Genetic Programming'', etc.-- that account for the
relevant journals and/or additional conferences. Papers with any of
these strings in its publication reference are directly classified
as EC papers and parsed as described above. If this criterion is not
fulfilled, then the title is scanned in order to detect another set
of relevant keywords such as ``evolutionary algorithm'', ``genetic
algorithm'', etc., or acronyms such as ``EA'', ``GA'' or ``GP''.
Again, if a paper triggers this criterion, it is classified as an EC
paper and processed accordingly. It must be noted that this system
has turned out to be rather accurate in detecting EC papers.
Actually, the visual inspection of the resulting network indicated
that only a small fraction of false positives (well below 1\% of the
total number of papers) passed the filters. These were mostly
computational biology papers, and were readily removed from the
network.

As a final consideration, we have chosen a large representative
sample of authors as the seed of our search robot. To be precise, we
have used a collection composed of all authors that have published
at least one paper in the last five years in any of the following
large EC conferences: GECCO, PPSN, EuroGP, EvoCOP, and EvoWorkshops
(unfortunately, CEC is not indexed in the DBLP; however, this does
not alter the macroscopic properties of the network, as it will be
shown below). This way, the immense majority of active EC
researchers is guaranteed to be included in the sample. Actually,
active authors not publishing in these fora are in practice linked
--directly or indirectly-- with all likelihood with authors who do
publish in them. Just as an indication, the number of authors used
as seed is 2,536 whereas the final number of authors in the network
is 5,492, that is, more than twice as many.

\section{Macroscopic Network Properties}
\label{sec:macro}
The overall characteristics of the EC co-authorship network are
shown in Table \ref{tab:todo} alongside with results obtained by
Newman \cite{newman01a}. The latter correspond to co-authorship
networks in Medline (biomedical research), the Physics E-print
Archive and SPIRES (several areas of physics and high-energy physics
respectively), and NCSTRL (several areas of computer science).

\begin{table}[t!]
\caption{Summary of results of the analysis of five scientific
collaboration networks.\label{tab:todo}}
\begin{center}
\begin{tabular}{lcccccc}
\hline
                            & EC     & Medline & Physics  & SPIRES & NCSTRL \\
                            \cline{2-6}
total papers                & 6199   & 2163923 & 98502    & 66652  & 13169  \\
total authors               & 5492   & 1520251 & 52909    & 56627  & 11994  \\
mean papers per author      & 2.9    & 6.4     & 5.1      & 11.6   & 2.6    \\
mean authors per paper      & 2.56   & 3.75    & 2.53     & 8.96   & 2.22   \\
collaborators per author    & 4.2    & 18.1    & 9.7      & 173.0  & 3.6    \\
size of the giant component & 3686   & 1395693 & 44337    & 49002  & 6396   \\
\ \ \ \ \ as a percentage   & 67.1\% & 92.6\%  & 85.4\%   & 88.7\% & 57.2\% \\
2nd largest component       & 36     & 49      & 18       & 69     & 42     \\
clustering coefficient      & 0.798  & 0.066   & 0.43     & 0.726  & 0.496  \\
mean distance               & 6.1    & 4.6     & 5.9      & 4.0    & 9.7    \\
diameter (maximum distance) & 18     & 24      & 20       & 19     & 31     \\
\hline
\end{tabular}
\end{center}
\end{table}

\begin{figure}[t!]
\centerline{\epsfxsize 6.5cm \epsfbox{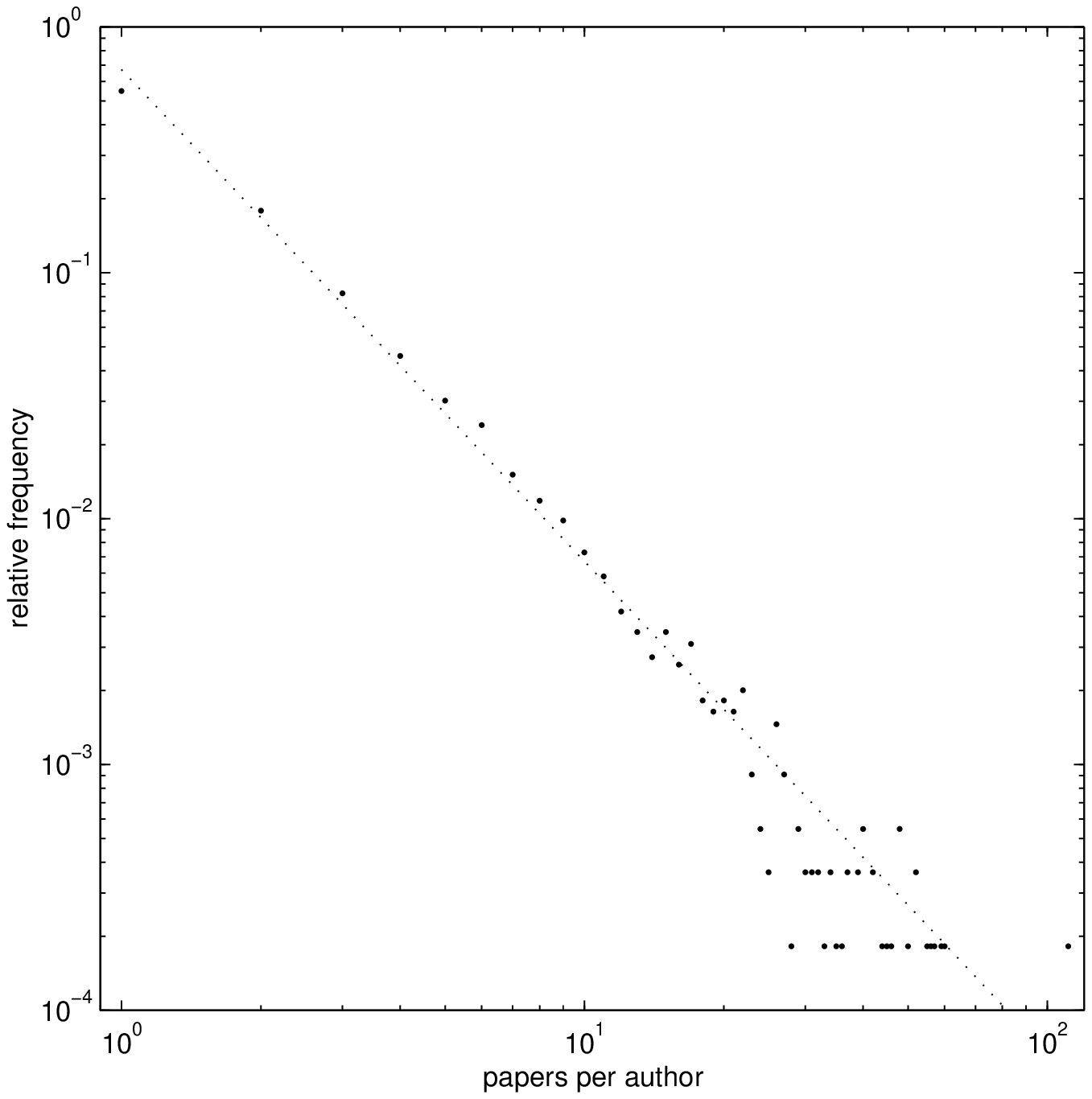}
\epsfxsize 6.5cm \epsfbox{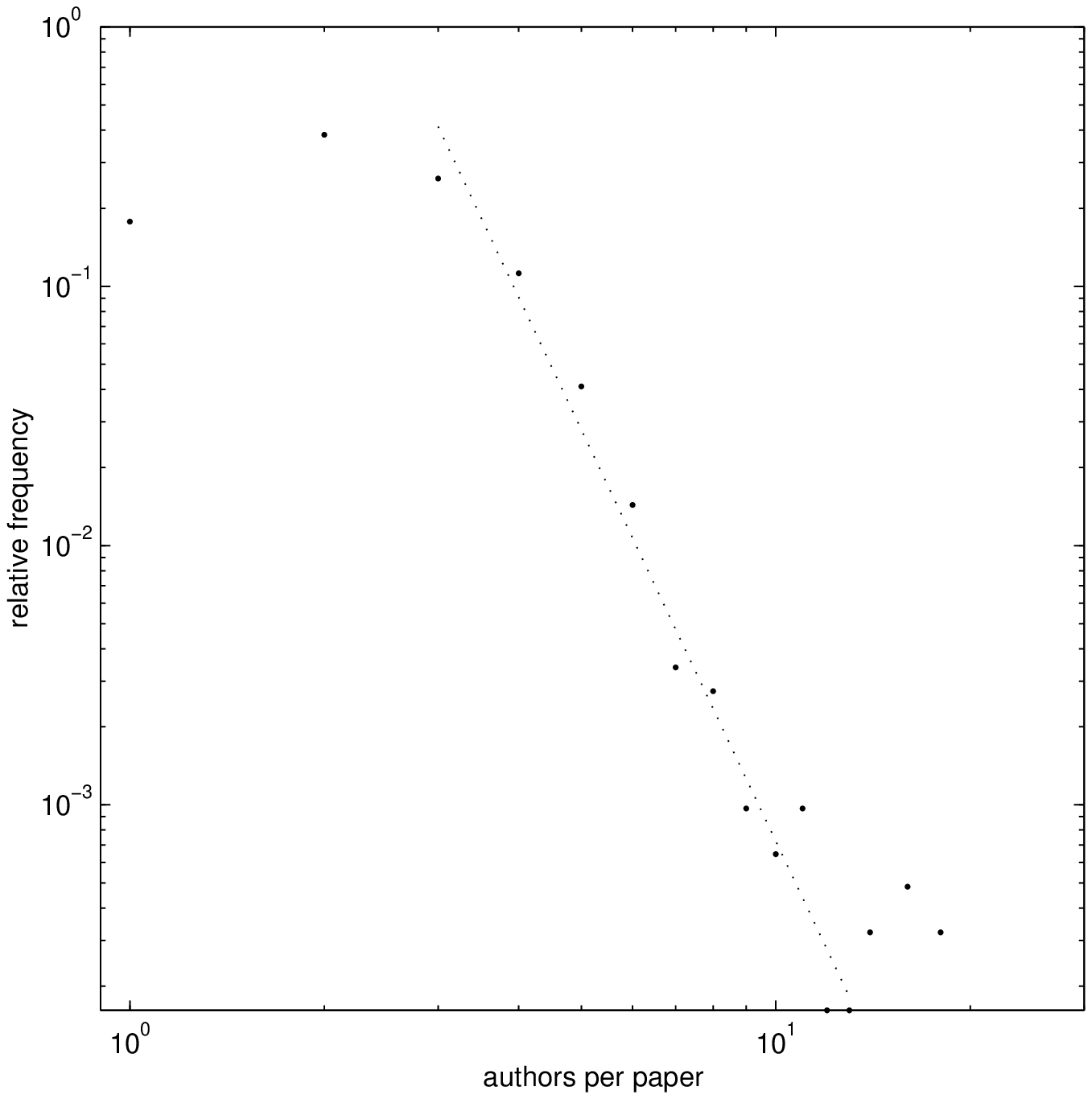}}
\caption{(Left) Histogram of the number of papers per author. The
slope of the dotted line is -2.00. (Right) Histogram of the number
of authors per paper. The slope of the dotted line is -5.27.}
\label{fig:fig1}
\end{figure}

First of all, the number of EC papers and authors is much smaller
than those for the communities studied by Newman; however, it must
be taken into account that these communities are much more general
and comprise different subareas. Notice also that in most aspects,
EC data seems closer to the NCSTRL database than to any other. This
indicates that despite the interdisciplinary nature of EC, the
publication practices of this area are in general those of computer
science. This way, average scientific productivity per author (2.9)
is not so high as in physics (5.1, 11.6) and biomedicine (6.4). It
nevertheless follows quite well Lotka's Law of Scientific
Productivity \cite{lotka26}, as shown by the power law distribution
illustrated in Fig. \ref{fig:fig1} (left). The most interesting
feature is the {\em long tail}: while most authors appear only once
in the database, there are quite a few that have authored dozens of
papers.

The average size of collaborations (2.56) is also smaller than in
biomedical research (3.75) or high-energy physics (8.96), although
similar to those of average physicists (2.53), and slightly superior
to average computer scientists (2.22). It also follows a power law
(up from 3 authors) as shown in Fig. \ref{fig:fig1} (right). Notice
the peak in the tail of the distribution, caused by the large
collaborations implied by proceedings. Their role will be examined
in Sect. \ref{sec:sociometric_stars}

Relevant considerations can be also done regarding the total number
of collaborators per author (4.2); physics and biomedicine are areas
in which new collaborations seem more likely than in EC (9.7, 173.0,
and 18.1). However, the figure for NCSTRL (3.6) is lower than for
EC, thus suggesting that the EC author is indeed open to new
collaborations, as regarded from a computer science perspective. The
histogram of number of collaborators per authors (not shown) also
fits quite well to a power law with exponent -2.58. In this case,
this power law can be attributed to a model of preferential
attachment such as the one proposed by Barab\'asi \cite{barabasi02}:
\emph{new} authors tend to link (be co-authors) of those that have
published extensively before. However, as we pointed out before,
that cannot be the whole story. For starters, information on who is
the most prolific author is not usually available (although educated
guesses can go a long way), and, besides, there are strong
constraints that avoid free linking: a person can only tutor so many
PhD students at the same time, for instance, and not everybody is
ready, or able, to move to the university of the professor she wants
to work with. However, let us point out that actors with many links
do not necessarily coincide with the most prolific; they are rather
persons that have diverse interests, reflected in their choice or
co-authors, participate in transnational projects, or have a certain
wanderlust, being visiting professors in many different
institutions, which leads them to co-author papers with their
sponsors or hosts in those institutions. The fact that the
clustering coefficient (that is, the average fraction of an actor's
collaborators that are collaborators themselves) in the EC
co-authorship networks is so high, and the mean degree of separation
is so close to the proverbial six degrees, means that in general all
authors in this field are no more than 6 degrees of separation of
those \emph{sociometric stars} with a wide variety of interests,
projects or visits. These sociometric stars will be analyzed more in
depth in next section.

\begin{figure}[t!]
\centerline{\epsfxsize 8cm \epsfbox{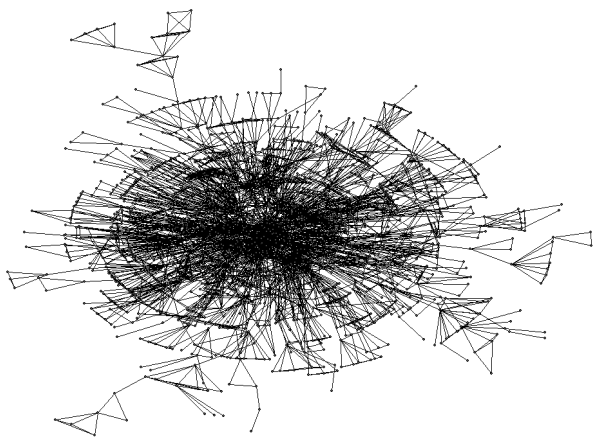}}
\caption{Graphical representation of the giant component of the EC
co-authorship network. A dense core with heavily connected authors
can be distinguished, with {\em tendrils} sprouting out of it that
include authors with less collaborators.} \label{fig:fig2}
\end{figure}

Another interesting aspect refers to the so-called \emph{giant
component}. This is a connected subset of vertices whose size
encompass most of the network. The remaining vertices group in
components of much smaller size (actually, independent of the total
size of the network). As pointed out in \cite{newman01a}, the
existence of this giant component is a healthy sign, for it shows
that most of the community is connected via collaboration, and hence
by person-to-person contact ultimately. In the case of the EC
network, the giant component comprises more than 2/3 of the network
(see Fig. \ref{fig:fig2}), again superior to the computer science
network, but significantly smaller than for physics or biomedicine.
This fact is nevertheless counteracted by the high clustering
coefficient (actually the highest of the set). This indicates a much
closer contact among actors, since one's collaborators are very
likely to collaborate among themselves too. It is also significant
that the mean distance among actors is halfway between the
medical/physics communities (around 4) and the computer science
community (around 9), while diameter is the second-smallest. This
shows that the EC community is halfway between computer science and
more theoretical fields, such as physics.

\section{Evolutionary Computation Sociometric Stars}
\label{sec:sociometric_stars}

In the previous section we have considered global collaboration
patterns that can be inferred from macroscopic properties of the
network. Let us know take a closer look at the fine detail of the
network structure. More precisely, we are going to identify which
actors play a more prominent role in the network, and analyze why
they are important. The term \emph{centrality} is used to denote
this prominency status for a certain node.

Centrality can be measured in multiple ways. We are going to focus
on metrics based on geodesics, i.e., the shortest paths between
actors in the network. These geodesics constitute a very interesting
source of information: the shortest path between two actors defines
a ``referral chain'' of intermediate scientists through whom contact
may be established -- cf. \cite{newman01b}. It also provides a
sequence of research topics (recall that common interests exist
between adjacent links of this chain, as defined by the co-authored
papers) that may suggest future joint works.

The first geodesic-based centrality measure that we are going to
analyze is \emph{betweenness} \cite{freeman77}, i.e., the total
number of geodesics between any two actors $i,j$ that passes through
a third actor $k$. The rationale behind this measure lies in the
information flow between actors: when a joint paper is written, the
authors exchange lots of information (research ideas, unpublished
results, etc.) which can in turn be transmitted (at least to some
extent) to their colleagues in other papers, and so on. Hence,
actors with high betweenness are in some sense ``hubs'' that control
this information flow; they are recipients --and emitters-- of huge
amounts of cutting-edge knowledge; furthermore, their removal from
the network would result in the increase of geodesic distances among
a large number of actors \cite{wassermanFaust94}.

The second centrality measure we are going to consider is precisely
based on this geodesic distance. Intuitively, the length of the
shortest path indicates the number of steps that research ideas (and
in general, all kind of memes) require to jump from one actor to
another. Hence, scientists whose average distance to other
scientists is small are likely to be the first to learn new
information, and information originating with them will reach others
quicker than information originating with other sources. Average
distance (i.e., \emph{closeness}) is thus a measure of centrality of
an actor in terms of their access to information.

\begin{table}[t!]
\caption{Most central actors in the EC network. D. E. Goldberg,
author of one of the most famous books on EC, figures prominently in
all rankings, as well as Kalyanmoy Deb, who is a well known author
in theoretical EC and multi-objective optimization. The rest of the
authors are well known as conference organizers, or as leaders of
some subfields within EC. The three columns show rankings for three
quantities: number of co-authors, and two centrality measures:
betweenness and closeness.}

\begin{center}
\begin{tabular}{llrllrllr}
\hline
     & \multicolumn{2}{c}{\# of co-workers} &\ \ &  \multicolumn{2}{c}{betweenness}    &\ \  &  \multicolumn{2}{c}{closeness}\\
     \cline{2-3}\cline{5-6}\cline{8-9}
1.  & K. Deb        & 98 && K. Deb         & 19.06 && K. Deb         & 28.60 \\
2.  & D.E. Goldberg & 75 && D.E. Goldberg  & 14.24 && W. Banzhaf     & 27.28 \\
3.  & R. Poli       & 67 && D. Corne       & 10.23 && D.E. Golberg   & 26.87 \\
4.  & M. Schoenauer & 62 && X. Yao         &  7.90 && R. Poli        & 26.86 \\
5.  & W. Banzhaf    & 58 && W. Banzhaf     &  7.70 && H.-G. Beyer    & 26.55 \\
6.  & D. Corne      & 56 && H. de Garis    &  6.92 && P.L. Lanzi     & 26.50 \\
7.  & X. Yao        & 56 && R. Poli        &  6.86 && D. Corne       & 25.93 \\
8.  & J.A. Foster   & 54 && J.J. Merelo    &  6.50 && M. Schoenauer  & 25.73 \\
9.  & J.J. Merelo   & 53 && H. Iba         &  6.48 && E.K. Burke     & 25.62 \\
10. & J.F. Miller   & 51 && M. Schoenauer  &  6.33 && D.B. Fogel     & 25.54 \\
\hline
\end{tabular}
\end{center}
\label{tab:centrality_1}
\end{table}

\begin{figure}[t!]
\centerline{\epsfxsize 6.5cm
\epsfbox{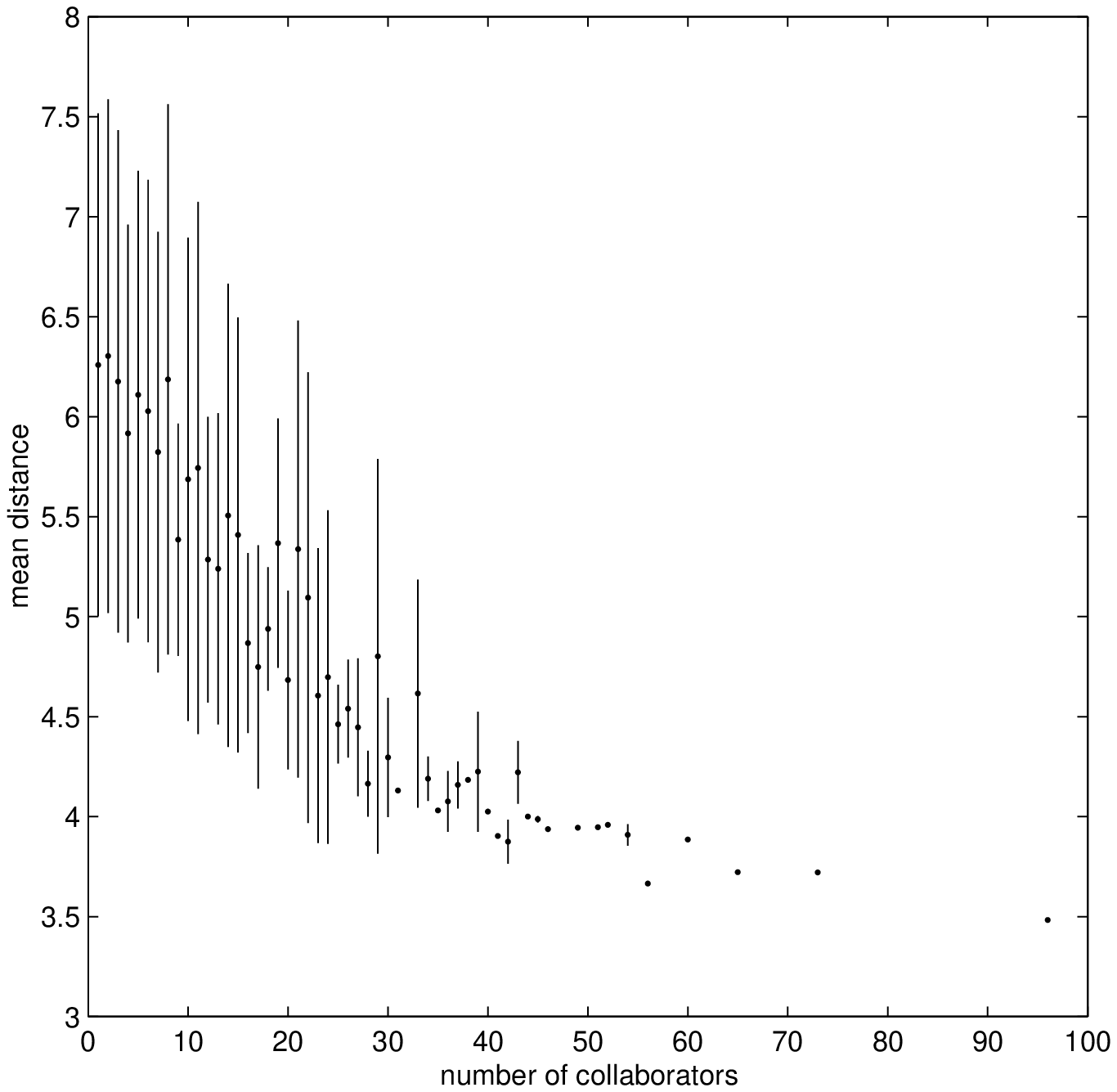} \epsfxsize
6.5cm \epsfbox{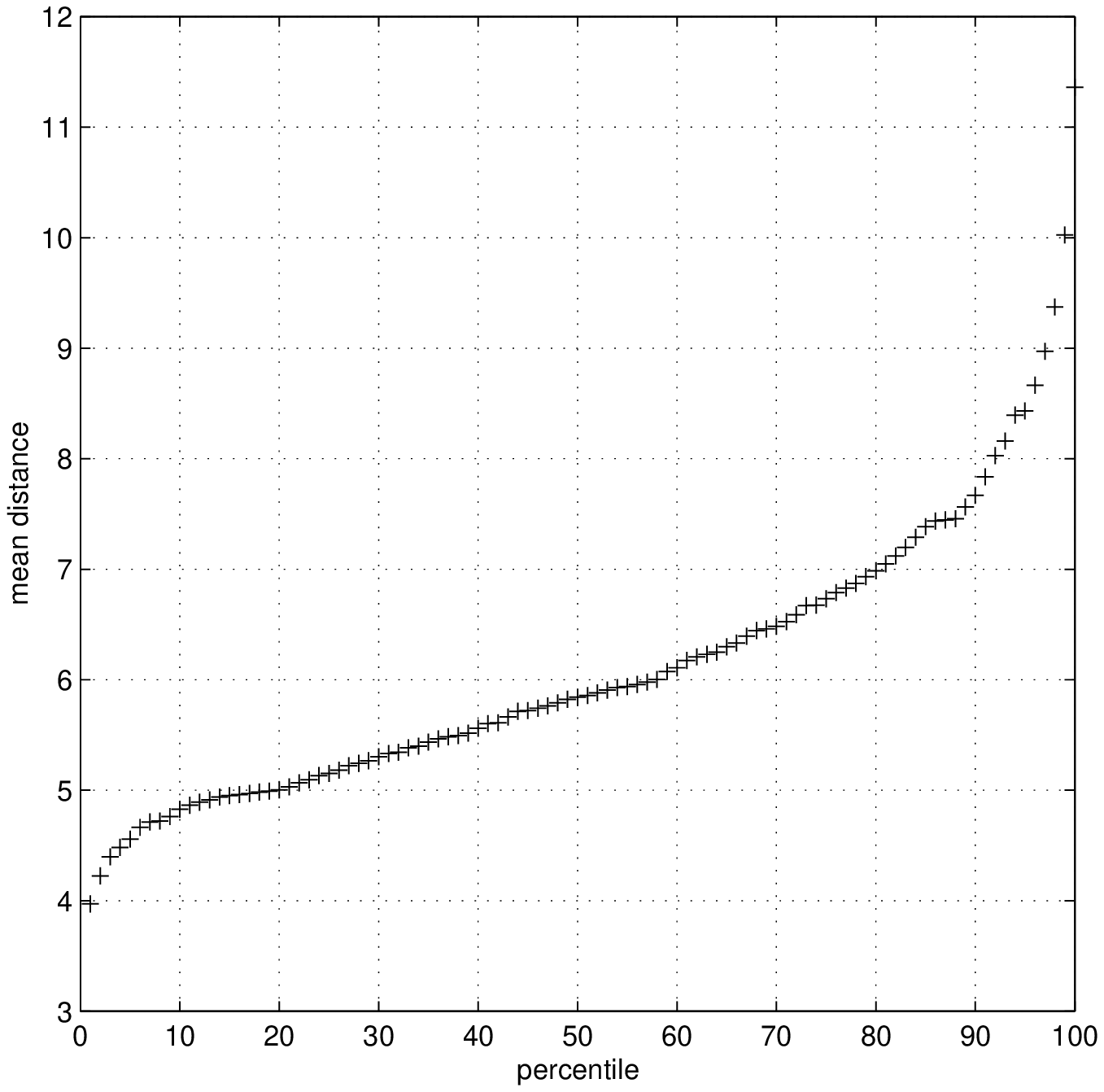}} \caption{(Left)
Mean distance to other authors as a function of the number of
collaborators. The error bars indicate standard deviations. (Right)
Percentile distribution of mean distances in the giant component.}
\label{fig:distance}
\end{figure}

The result of our centrality analysis of the EC network is shown in
Table \ref{tab:centrality_1}. The numbers provided for each actor
indicate the normalized values of betweenness and closeness (that
is, their actual values divided by the maximum possible value,
expressed as a percentage). Regarding betweenness, the analysis
provides clear winners, with large numerical differences among the
top actors. These differences are not so marked for closeness values
with all top actors clustered in a short interval. Notice that there
are some actors that appear in both top-lists. Using Milgram's
terminology \cite{milgram67}, these constitute the \emph{sociometric
superstars} of the EC field.

Several factors are responsible for the prominent status of these
actors. Obviously, scientific excellence is one of them. This
excellence is difficult to measure in absolute, objective terms, but
the number of collaborators provides some hints on it\footnote{This
quantity is strongly correlated with the number of papers
($\rho=.82$), and thus provides information on the efficiency in
knowledge transmission, which is the ultimate goal of scientific
publishing. Involvement in PhD supervision and research projects,
and wide research interests will typically result in a higher number
of collaborators as well.}. This quantity is shown for the top ten
actors in the network in Table \ref{tab:centrality_1}. Certainly,
some correlation between degree and centrality is evident. This is
further illustrated in Fig. \ref{fig:distance} (left). As it can be
seen, there is a trend of decreasing average distance to other
actors as the actor degree increases. By crossing this information
with the percentile distribution of distances shown in Fig.
\ref{fig:distance} (right) we can obtain some interesting facts
about the collaborative strength of elite scientists. For example,
consider the top 5\% percentile; it is composed of actors whose
average distance to the remaining actors is at most 4.61. According
to Fig. \ref{fig:distance} (left), 23 collaborators are required at
least to have an average distance below this value. A more sensitive
analysis indicates that 33 collaborators are required to have an
statistically significant (using a standard t-test) result.

\begin{figure}
\centerline{\epsfxsize 7.2cm \epsfbox{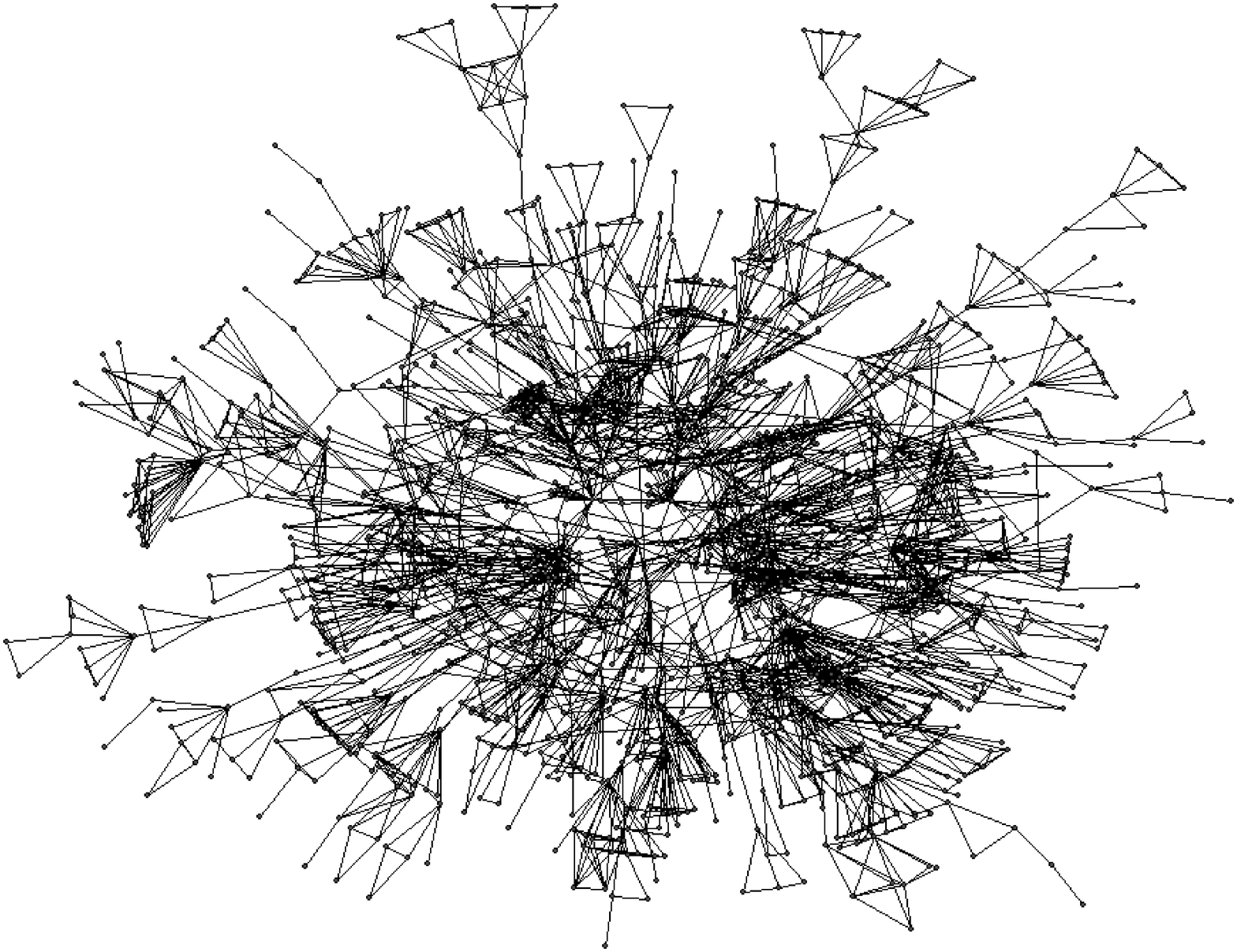}
\epsfxsize 6cm \epsfbox{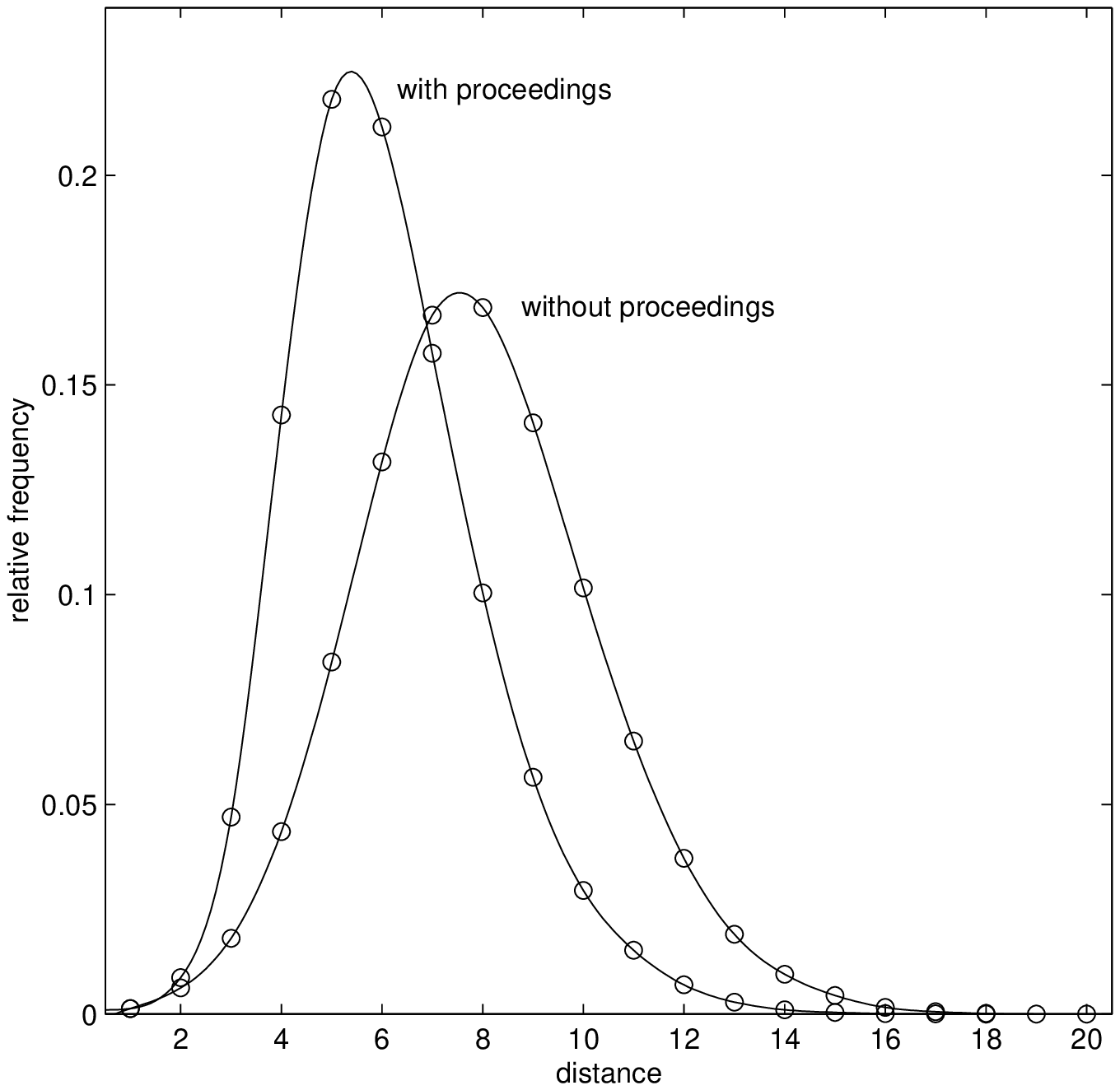}} \caption{(Left)
Graphical representation of the network after removing proceedings.
(Right) Comparison of the distribution of author distances with and
without proceedings. The solid lines are eye-guides.}
\label{fig:noproc}
\end{figure}

Another important factor influencing the particular ranking shown
above is the presence of conference proceedings among authors'
publications. These play a central role in the creation and
structure of the network, to the point that its features change
dramatically if links arising from proceedings co-authorship are
removed. To begin with, the visual aspect of the network is
different, as is shown in the left hand side of Fig.
\ref{fig:noproc} (compare it to the network with proceedings
included, shown in Fig. \ref{fig:fig2}). The reader should notice
that the core is much more diffuse (actually, it looks like there
are several micro-cores, plausibly corresponding to different EC
subareas).

This change is also reflected in the right hand side of Fig.
\ref{fig:noproc}, which plots the histogram of average distances
from each node to the rest of the network: without proceedings, the
average distance and maximum distance increase by 2 units, and the
modal distance increases by 3 units. The resulting distribution is
also much more symmetric than the original distribution, which was
notably skewed towards low values. This can be explained by the very
distinctive \emph{authoring} (in property, \emph{editing}) patterns
of proceedings: they are usually edited by a larger number of
researchers, typically corresponding to the different thematic areas
included in the conference or symposium. These are often senior
researchers, with a prominent position in their subareas (thus,
centrality and proceeding editorship reinforce each other).
Furthermore, the fact that editors come from different areas
contribute to the creation of long-distance links, resulting in a
dramatic overall decrease of inter-actor distances.

\begin{table}[t]
\caption{Most central actors in the EC network after removing
proceedings.}
\begin{tabular}{llrllrllr}
\hline
     & \multicolumn{2}{c}{\# of co-workers} &\ \ &  \multicolumn{2}{c}{betweenness}    &\ \  &  \multicolumn{2}{c}{closeness}\\
     \cline{2-3}\cline{5-6}\cline{8-9}
1.  & D.E. Goldberg  & 63 && D.E. Goldberg  & 22.68              & & Z. Michalewicz & 20.21 \\
2.  & K. Deb         & 55 && K. Deb         & 20.04              & & K. Deb         & 20.05 \\
3.  & M. Schoenauer  & 52 && M. Schoenauer  & 12.68              & & M. Schoenauer  & 19.89 \\
4.  & X. Yao         & 42 && H. de Garis    & 12.62              & & A.E. Eiben     & 19.77 \\
5.  & H. de Garis    & 41 && Z. Michalewicz & 12.58              & & B. Paechter    & 19.70 \\
6.  & T. Higuchi     & 40 && T. B\"ack      & 10.31              & & D.E. Goldberg  & 19.64 \\
7.  & Z. Michalewicz & 40 && R.E. Smith     &  9.46              & & T. B\"ack      & 18.70 \\
8.  & L.D. Whitley   & 39 && X. Yao         &  9.07              & & D.B. Fogel     & 18.59 \\
9.  & M. Dorigo      & 38 && A.E. Eiben     &  8.61              & & J.J. Merelo    & 18.52 \\
10. & J.J. Merelo    & 38 && B. Paechter    &  8.05              & & T.C. Fogarty   & 18.50 \\
\hline
\end{tabular}
\label{tab:noprocs}
\end{table}

Although proceeding editorship is certainly a scientific activity,
and constitutes a valuable contribution to the community, putting
them at the same level of research papers is arguable at the very
least. It thus seems reasonable to exclude proceedings from the
network to obtain a more unbiased figure of centrality. We have done
this, obtaining the results shown in Table \ref{tab:noprocs}. As it
can be seen, there is now a higher agreement between the two
centrality measures (7/10 are the same, vs. 6/10 before).
Furthermore, researchers of unquestionable scientific excellence who
were not in the previous ranking do appear now. For example, Z.
Michalewicz, author of several excellent EC books, is now the author
with the highest closeness, the 5th-highest betweeness, and the
7th-highest number of collaborators. Overall, this may provide a
more objective view on the central actors of our field.

\section{Discussion and Conclusion}

In this paper, we have made a preliminary study of the co-authorship
network in the field of evolutionary computation, paving the way to
study the impact of certain measures, such as grants, the
establishment of scientific societies or new conferences, has on the
subject. The general features of the network suggest that it is
quite similar to the field it can be better placed, computer
science, but, at the same time, authors are much more closely
related with each other. We have also taken into account the impact
co-editorship of proceedings have on the overall aspect of the
network and most centrality measures. To the best of our knowledge,
this issue had not been considered in previous related works, and we
believe it plays an important role in distorting some network
properties. We suggest to not consider them in the future in this
kind of studies.

In connection to this latter issue, we believe that co-authorship
networks created by different kind of papers (technical reports,
conference papers, journal papers) might be different owing to the
different kind of collaboration they imply. Consider that while
technical reports may be written in a hurry and present very
preliminary results, conference papers are usually somewhat more
long term, and journal papers really indicate a committed scientific
relationship (due to the long time they take to be published and the
several iterations of the revision process). The authors suggest to
approach them separately and analyze the features of the networks
they yield.

In addition to this, our future lines of work along this topic will
include the analysis of the network evolution through time, as well
as the impact funded scientific networks and transnational grants
(such as EU grants) have had on it. We also plan to study the
existence of \emph{invisible colleges} or communities within the EC
field, and analyze which their axes of development are, e.g.,
topical or regional.

\bibliographystyle{apalike}
\bibliography{coauthorship}

\end{document}